\documentclass[preprint,showpacs,showkeys,preprintnumbers,amsmath,amssymb,PRL,endfloats,floatfix]{revtex4}

\usepackage{graphicx}
\usepackage{dcolumn}
\usepackage{bm}

\begin{document}

\preprint{ }

\title{Phosphorus donors in highly strained silicon}

\author{Hans Huebl}
\email[corresponding author ]{huebl@wsi.tum.de}
\affiliation{Walter Schottky Institut, Technische Universit\"{a}t
M\"{u}nchen, Am Coulombwall 3, 85748 Garching, Germany}
\author{Andre R. Stegner}
\affiliation{Walter Schottky Institut, Technische Universit\"{a}t
M\"{u}nchen, Am Coulombwall 3, 85748 Garching, Germany}
\author{Martin Stutzmann}
\affiliation{Walter Schottky Institut, Technische Universit\"{a}t
M\"{u}nchen, Am Coulombwall 3, 85748 Garching, Germany}
\author{Martin S.~Brandt}
\affiliation{Walter Schottky Institut, Technische Universit\"{a}t
M\"{u}nchen, Am Coulombwall 3, 85748 Garching, Germany}
\author{Guenther Vogg}
\affiliation{Fraunhofer Institut f{\"u}r Zuverl{\"a}ssigkeit und
Mikrointegration IZM, Institutsteil M{\"u}nchen, Hansastrasse 27d,
80686 M\"{u}nchen, Germany}
\author{Frank Bensch}
\affiliation{Fraunhofer Institut f{\"u}r Zuverl{\"a}ssigkeit und
Mikrointegration IZM, Institutsteil M{\"u}nchen, Hansastrasse 27d,
80686 M\"{u}nchen, Germany}
\author{Eva Rauls}
\affiliation{Aarhus Universitet, Institut for Fysik og Astronomi,
Ny Munkegade, Bygn.~1520, 8000 Aarhus C, Denmark}
\author{Uwe Gerstmann}
\affiliation{Institut de Min\'eralogie et de Physique des Milieux Condens\'es,
Universit\'e Pierre et Marie Curie,
Campus Boucicaut, 140 rue de Lourmel, 75015 Paris, France}

\date{\today}

\begin{abstract}
The hyperfine interaction of phosphorus donors in fully strained
Si thin films grown on virtual Si$_{1-x}$Ge$_x$ substrates with
$x\leq 0.3$ is determined via electrically detected magnetic
resonance. For highly strained epilayers, hyperfine interactions
as low as 0.8~mT are observed, significantly below the limit
predicted by valley repopulation. Within a Green's function
approach, density functional theory (DFT) shows that the
additional reduction is caused by the volume increase of the unit
cell and a local relaxation of the Si ligands of the P
donor.\end{abstract}


\maketitle


At present, several approaches for solid-state based quantum
computing hardware are actively pursued. The possible integration
with existing microelectronics and the long decoherence times
\cite{castner62, tyryshkin03, gordon58} are particular advantages
of concepts \cite{kane98, kane00,vrijen00, hollenberg04} using the
nuclear or electronic spins of phosphorus donors in group IV
semiconductors as qubits. These concepts require gate-controlled
exchange coupling between neighbouring donors. However, to control
the exchange coupling in semiconductors with an indirect band
structure such as Si, the donor atoms have to be positioned with
atomic precision~\cite{schofield03} due to Kohn-Luttinger
oscillations of the donor wavefunction~\cite{kane00, koiller02,
wellard03}. Under uniaxial compressive strain in [001] direction,
two conduction band (CB) minima are lowered in energy, which leads
to a suppression of the oscillatory behaviour of  the wavefunction
for donors located in the (001) lattice plane \cite{koiller02}.
The strain will also affect the wavefunction at the position of
the donor atom, which can be observed directly via the hyperfine
(hf) interaction between the donor electron and its nucleus
\cite{wilson61, koiller02}.
In this letter, we experimentally and theoretically study the hf
interaction of phosphorus donors in silicon as a function of
uniaxial compressive strain in thin layers of Si on virtual SiGe
substrates, extending the regime investigated by Wilson and Feher
\cite{wilson61} by a factor of 20 to higher strains. We find that
the reduction of the hf interaction significantly exceeds the
limit predicted so far \cite{wilson61, koiller02, kane00}. By {\em
ab-initio} DFT calculations using a Green's function approach
\cite{overhof04}, we are able to show that this reduction is
caused by the strain-induced lattice distortion, and a relaxation
of the Si lattice next to the P atoms. As the nuclear exchange
mediated by the overlap of the donor wavefunction scales with the
square of the hf interaction, these results have a direct impact
on the rate at with two-qubit operations can be performed
\cite{kane98}.

\begin{figure}
\includegraphics[width=10cm]{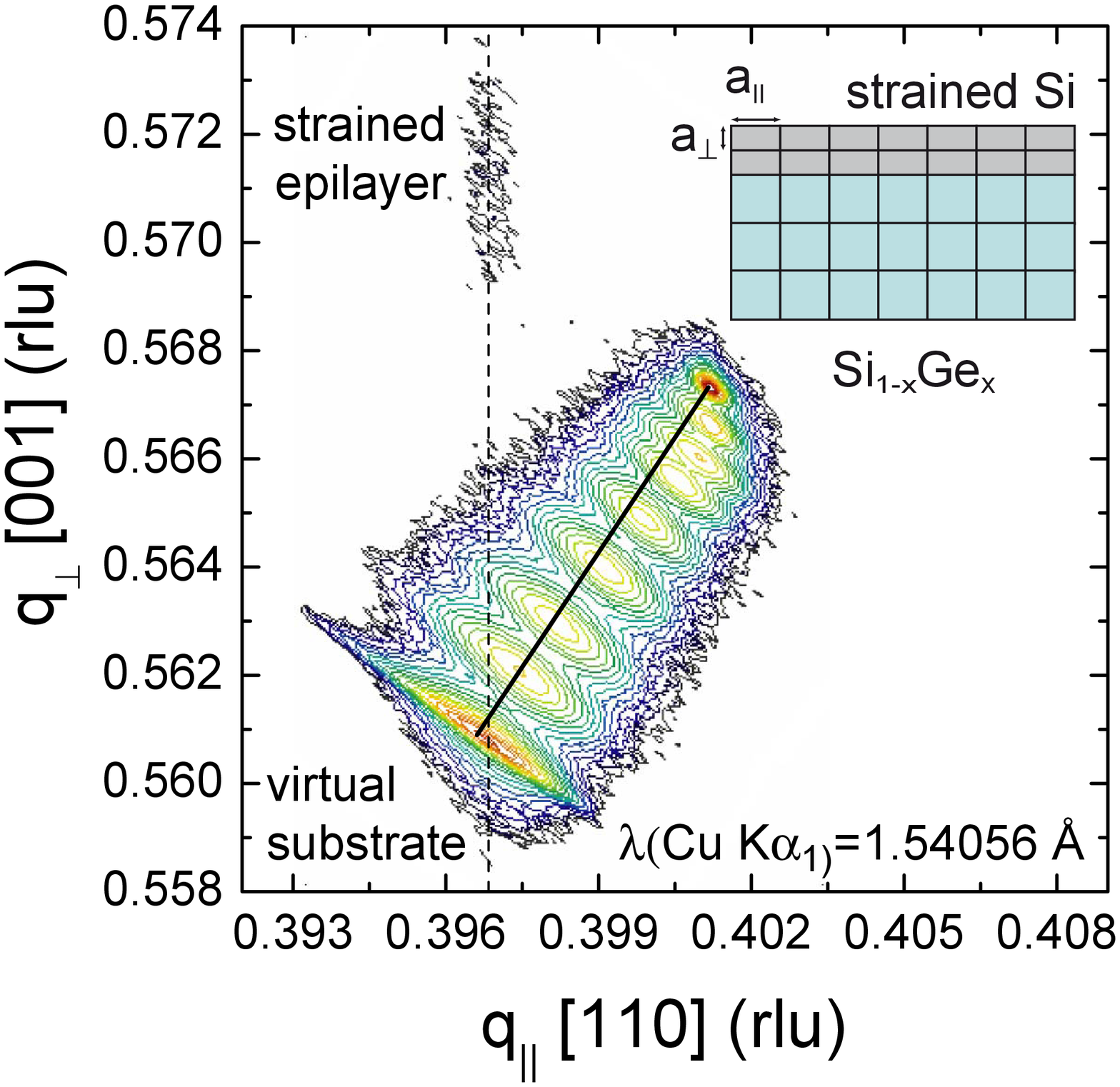}
\vspace*{-0.3cm} \caption{\label{fig:RSM} (color online)
Reciprocal space map of the 224 reflex of a 15~nm thick fully
strained Si film on a Si$_{0.7}$Ge$_{0.3}$ virtual substrate
(log.-scaled iso-intensity plot, Cu~$K\alpha_1$ radiation). The
straight line represents $q_{\parallel}$ and $q_{\perp}$ predicted
for relaxed Si$_{1-x}$Ge$_{x}$ with $0\leq x \leq 0.3$
\cite{becker82}.} \vspace*{-0.2cm}
\end{figure}

The samples under study were prepared by chemical vapor deposition
(CVD) on $30~\rm{\Omega cm}$ B-doped Si (001) substrates
\cite{kreuzer05}. Fully strained thin P-doped Si epilayers with
$[P]\simeq 1\times 10^{17}~\rm{cm^{-3}}$ were grown lattice
matched on virtual relaxed Si$_{1-x}$Ge$_x$-substrates optimized
for low dislocation densities. These substrates consist of an
intrinsic 0.3~$\rm{\mu m}$ Si buffer, a variable sequence of
Si$_{1-y}$Ge$_y$ buffer layers with stepwise increasing $y$, and
the actual 2~$\rm{\mu m}$ thick virtual Si$_{1-x}$Ge$_x$ substrate
with Ge-contents $x=0.07, 0.15, 0.20, 0.25,$ and $0.30$.
High-resolution X-ray diffraction (HRXRD) on the 004 and 224
reflexes was used to determine the (in- and out-of-plane) lattice
constants of the SiGe layers from which the Ge content and degree
of relaxation were calculated using a parabolic dependency of $x$
on the relaxed lattice constants \cite{dismukes64} and a linear
interpolation of the elastic stiffness constants between Si and Ge
\cite{LBSielasticmoduli01}. The Si$_{1-x}$Ge$_x$ layer determines
the strain of the Si epilayer: The larger lattice constant of SiGe
alloys compared to Si leads to biaxial tensile strain, accompanied
by a compensating uniaxial compressive strain in growth direction
as shown by the inset in Fig.~\ref{fig:RSM}. All Si:P epilayers
have a thickness of 15~nm, well below the critical thickness of Si
on Si$_{0.7}$Ge$_{0.3}$ \cite{samavedam99}.

Figure~\ref{fig:RSM} shows the corresponding HRXRD reciprocal
space-map (RSM) of the 224 reflex for the Si$_{0.7}$Ge$_{0.3}$
sample. The peak arising from the Si wafer is set to match the
lattice constant of c-Si ($5.4310$~{\AA})~\cite{becker82}. Extending
to lower reciprocal lattice units
$q_{\parallel}$=$\sqrt{2}\lambda/a_{\parallel}$ (in plane) and
$q_{\perp}$=$\sqrt{4}\lambda/a_{\perp}$ (out of plane), a set of
diffraction peaks arising from the SiGe buffer layers is observed,
followed by the peak of the virtual substrate. The slight
difference between the observed values and those expected for
fully relaxed Si$_{1-x}$Ge$_{x}$ alloys (black solid
line,~\cite{dismukes64}) indicates a degree of relaxation of
96.4\% of the virtual substrate.
%
%
Additionally, Fig.~\ref{fig:RSM} shows at $q_{\perp}$=$0.5715$
($a_{\perp}$=$5.3912$~{\AA}) and $q_{\parallel}$=$0.39686$
($a_{\parallel}$=$5.4898$~{\AA}) the diffraction by the strained
silicon epilayer. The perfect agreement of $q_\parallel$ of the Si
epilayer and the virtual substrate reflects the fully strained
pseudomorphic growth.
It is important to note that the compression in growth direction
is in accordance with linear elasticity theory. With
$C_{11}=165.6$~GPa and $C_{12}=63.9$~GPa as the stiffness
constants of Si \cite{LBSielasticmoduli01}, the relation
$\epsilon_\perp$=$-2C_{12}/C_{11}\cdot\epsilon_\parallel$ between
in-plane and out-of-plane strains
$\epsilon_\parallel$=$(a_{\parallel}$--$a_{{\rm{Si}}})/a_{{\rm{Si}}}$
and $\epsilon_\perp$=$(a_{\perp}$--$a_{{\rm{Si}}})/a_{{\rm{Si}}}$
predicts a value of $a_\perp$=$5.3856$~{\AA} on the
Si$_{0.7}$Ge$_{0.3}$ substrate, in reasonably good agreement with
$a_\perp$ obtained from Fig.~\ref{fig:RSM}, taking into account
the width of the thin Si layer caused by the small thickness.
Hence, the strain values for the different strained Si samples are
obtained from the relaxed in-plane lattice constants of the
corresponding SiGe buffers using linear elasticity theory.

To observe the the small amount of P donors in the thin strained
layer with high sensitivity, we use electrically detected magnetic
resonance (EDMR), which monitors spin resonance via the influence
of spin selection rules on charge transport processes
\cite{schmidt66, stich96, brandtpss04}. The EMDR experiments were
performed in a dielectric ring microwave resonator using a
HP83640A microwave source and measuring the spin-dependent
photoconductivity  at $T=5$~K in a liquid He flow cryostat under
illumination with white light from a tungsten lamp. Resonant
changes $\Delta I/I\approx10^{-5}$ of the photocurrent $I$ were
detected using a current amplifier, lock-in detection and magnetic
field modulation with 0.2~mT at 1.124~kHz. All spectra were
obtained using a microwave power of 50~mW and are normalized to a
microwave frequency of 9.749~GHz.

\begin{figure}
\includegraphics[width=10cm]{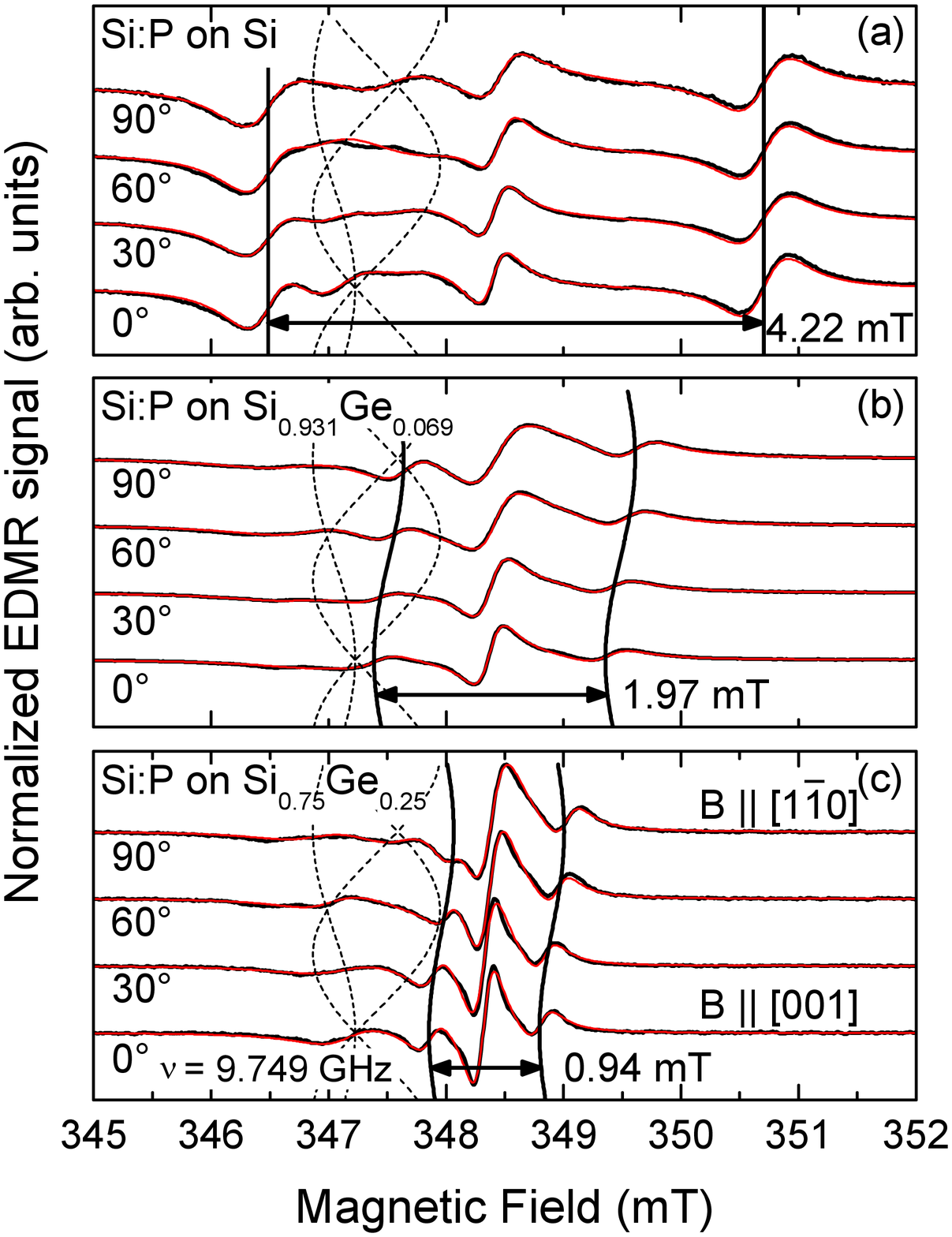}
\vspace*{-0.3cm} \caption{\label{fig:spec} (color online)
Electrically detected magnetic resonance of P donors in relaxed
and fully strained Si epilayers on three different
Si$_{1-x}$Ge$_x$ virtual substrates as a function of the
orientation of the [001] growth direction with respect to the
magnetic field $\vec{B}_0$ for a rotation around $[110]$. To
extract spectroscopic information, the spectra have been fitted
using Lorentzian lines. The sum of these lines is shown in red. }
\vspace*{-0.2cm}
\end{figure}

Figure~\ref{fig:spec} shows the EDMR signal observed as a function
of the applied magnetic field for different angles between the
[001] direction of the Si epilayer and the externally applied
magnetic field $\vec{B}_0$ (rotation around $[110]$). To extract
the spectroscopic information, the spectra were fitted with
Lorentzian lines. In the range of $B_0=346.5$ to 348~mT,
anisotropic lines due to the P$_{\rm{b0}}$ center at the
Si/SiO$_2$ interface are observed with their characteristic
anisotropy~\cite{poindexter81, stesmans98} (dashed lines in
Fig.~\ref{fig:spec}). Within the resolution of our experiment,
these resonances remain unchanged in the strained layers. The
dominant central line with $g=1.9994$ in the unstrained sample
becomes slightly anisotropic under strain and could be assigned to
CB electrons \cite{young97}. In the unstrained Si layer (cf.
Fig.~\ref{fig:spec}~(a)), also the characteristic two hf-split
lines of P in Si with a separation of $A_{\rm HF}=4.2$~mT are
easily resolved. The hf lines are isotropic as indicated by the
black vertical lines with a center of gravity at $g$=$1.9985$,
well known for unstrained c-Si \cite{young97, cullis75}.

In contrast, the hf-split resonances become anisotropic for a
strained epilayer as indicated by the black lines in
Fig.~\ref{fig:spec} (b) and (c). For Si on a
Si$_{0.93}$Ge$_{0.07}$ substrate $(\epsilon_\perp$=$-0.00199)$,
they can be described with an isotropic hf interaction of 1.97~mT
and an anisotropic $g$-factor which varies by $\Delta
g$=$(1.46\pm0.1)\times 10^{-3}$ from $\vec{B}_0 \parallel
[1\bar{1}0]$ to $\vec{B}_0 \parallel [001]$. Figure~\ref{fig:spec}
(c) shows the results for the epilayer with
$\epsilon_\perp$=$-0.00729$ obtained on a Si$_{0.75}$Ge$_{0.25}$
substrate. Here, the isotropic hf splitting decreases to 0.94~mT,
while $\Delta g$=$(1.21\pm 0.06)\times 10^{-3}$. The small
P-related hf splittings exclude significant segregation of P into
the SiGe layers during the growth process, which would lead to
larger splittings \cite{vollmer74, hoehne95}.

\begin{figure}
\includegraphics[width=10cm]{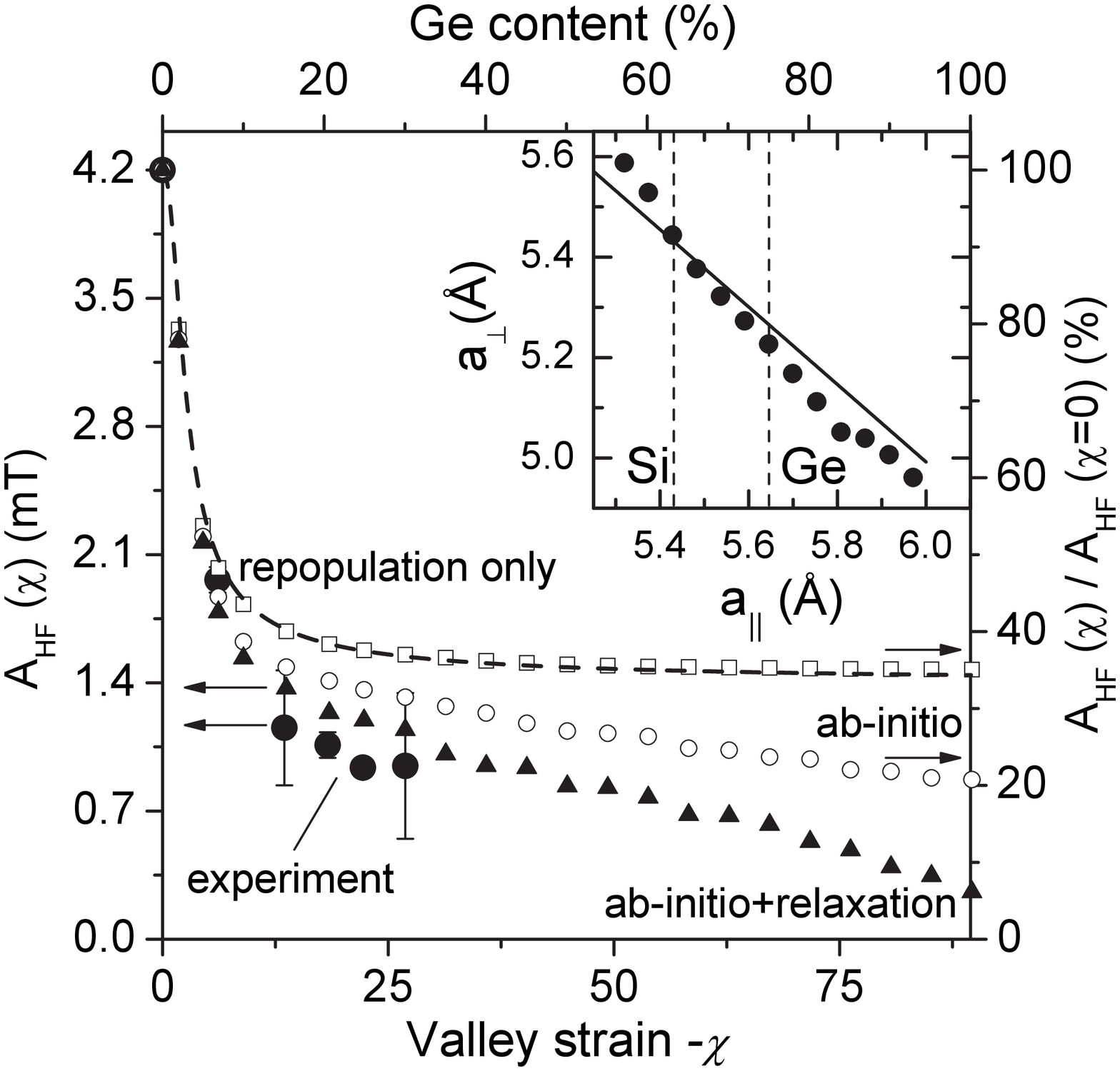}
\vspace*{-0.3cm} \caption{\label{fig:sum}Hyperfine splittings
observed for P in fully strained epilayers on Si$_{1-x}$Ge$_x$
substrates as a function $x$ and the resulting valley strain. The
black dots indicate the experimental data. The dashed line
represents the behaviour expected by valley repopulation
\cite{wilson61}. The DFT results for valley repopulation only are
indicated by open squares, the results including the
strain-induced change in the volume of the unit cell by open
circles, and those additionally including relaxation of the P
nearest neighbors by filled triangles. For the latter and the
experimental data the absolute values $A_{\rm HF}$ are shown, the
remaining data refers to the relative scale.
The insert shows the out-of-plane lattice constant $a_\perp$ of
thin Si layers (thickness 12 unit cells) as a function of the
in-plane lattice constant $a_\parallel$ predicted by SCC-DFTB
compared to linear elasticity theory indicated by the straight
line. Pseudomorphic Si layers on pure Si and Ge substrates are
shown by dashed lines.} \vspace*{-0.2cm}
\end{figure}

For an isolated P donor atom in unstrained Si, the isotropic
Fermi-contact hf interaction $A_{\rm HF}$ is given by $8/3 \mu_B
\pi |\psi(0)|^2$ \cite{wilson61, KohnSSP57}, where $\mu_B$ is
Bohr's magneton and $|\psi(0)|^2$ is the probability amplitude of
the unpaired electron wave function at the nucleus, giving rise to
the doublet of hf lines separated by $A_{\rm HF}=4.2$~mT in
Fig.~\ref{fig:spec}(a). To determine $|\psi(0)|^2$, we first note
that the cubic crystal field leads to the formation of a singlet
ground state plus a doublet and a triplet of excited states
instead of a six-fold degenerate ground state of the donor in Si.
Only the fully symmetric singlet ground state has a non-vanishing
probability amplitude at the nucleus, so that $\psi$ can be
written as a superposition $\psi=\sum_{i=1}^6 (1/\sqrt{6}) \Phi_i$
of the six valleys contributing to the donor. Here, each $\Phi_i$
is a product of the corresponding CB Bloch wavefunction and a
hydrogenic envelope-function. The probability of the unpaired
electron  at the nucleus $|\psi(0)|^2$ becomes $1/6|\sum_{j=1}^6
\Phi_i(0)|^2$=$6|\Phi(0)|^2$, since due to symmetry
$\Phi_i(0)$=$\Phi(0)$ for all $i$. Assuming that the only effect
of strain is the change of relative population of the CB minima,
we similarly find $\psi$=$\sum_{i=1}^2 (1/\sqrt{2}) \Phi_i$ and
$|\psi(0)|^2$=$2|\Phi(0)|^2$ under high uniaxial strain, when only
two CB minima contribute. Therefore, in the fully strained case,
the hf interaction should be 1/3 of the unstrained case. In
contrast, in Fig.~\ref{fig:spec} we already observe a reduction to
$0.21 \cdot A_{\rm{HF}}$, clearly below the $0.33\cdot
A_{\rm{HF}}$ limit obtained above. Based on group and linear
elasticity theory, Wilson and Feher \cite{wilson61} have evaluated
the analytical dependence of $A_{\rm{ HF}}(\chi)$ on the so-called
valley strain
$\displaystyle{\chi=-\frac{\Xi_u}{3 \Delta_c}\left(1+\frac{2\;
C_{12}}{C_{11}}\right) \epsilon_{\parallel} }$,
where $\Xi_u=8.6$~eV \cite{tekippe72} is the uniaxial deformation
potential, and  $6\Delta_c=2.16$~meV~\cite{ramdas81} is the energy
splitting of the singlet and doublet state in the unstrained
material. In Fig.~\ref{fig:sum}, a comparison of the prediction of
Eq.~(2) in Ref.~\cite{wilson61}(dashed line) with the hf
splittings determined experimentally (full circles) clearly shows,
that pure valley repopulation is not able to describe the
experimental data for $x>0.07$. An empirical treatment of
additional radial redistribution effects as discussed in
Ref.~\cite{fritzsche62} would lead to $0.30\cdot A_{\rm HF}$ for
$\chi=-89$ ($x=1$), only a slight reduction of the repopulation
limit and, thus, still at strong variance with the experimental
data.

An {\em ab-initio} calculation of hf interactions is necessary to
clarify the situation, but is a demanding task. Already for the
unstrained case, the delocalisation of the donor wave function
leads to problems in describing the magnetisation distribution
correctly~\cite{hale69}. Whereas the energetic position of shallow
levels as well as the geometries obtained by usual LSDA supercell
calculations can be expected to be in accordance with experimental
data, the hf interactions are generally
overestimated~\cite{overhof04,rauls04}. Recently, a Green's
functions approach (LMTO-GF) has been shown to circumvent this
problem of usual supercell approaches~\cite{overhof04}. By
carefully analyzing the donor-induced resonance at the bottom of
the CB, it becomes possible to describe the central-cell
correction to effective mass theory by first principles with an
accuracy that allows a prediction of superhyperfine interactions
for shallow donor states including the Kohn-Luttinger
oscillations~\cite{kohn55,overhof04}.

In the case of a strained host material, however, the situation
becomes more complicated, since excited states are admixed to the
former pure singlet ground state. An application of density
functional theory (DFT) is only possible in combination with
linear elasticity theory:
Due to the applied strain, the symmetry of the P donor is reduced,
and the resonance at the bottom of the CB is transforming
according the $a_1$ representation in the unstrained case, now
shows admixtures of the $b_1$ and $b_2$ representations of
$D_{2d}$ symmetry. The location of the P donor atom in their nodal
planes implies a correlation of these $b_1$ and $b_2$-like
orbitals with the admixed doublet state. Since, furthermore, only
one component of the diamagnetic doublet state contributes to the
singlet ground state under strain~\cite{wilson61}, it is
reasonable to construct the spin-densities, which enter the
self-consistent LSDA total energy calculations for a given valley
strain, by
$n^{\sigma}(r)$=$(1-\alpha(\chi))\cdot n^{\sigma}_{a1}(r)+
\alpha(\chi)\cdot n^{\sigma}_{b1}(r)$,
where $\sigma=\mid\uparrow\rangle$ or $\mid\downarrow\rangle$.
Here, $\alpha(\chi)$ is obtained from the strain-dependent
admixture of the doublet states determined by linear elasticity
theory (see Eq.~(C6) in Ref.~\cite{wilson61}).
Figure~\ref{fig:sum} shows that the spin densities constructed
this way allow a reasonable description of the pure valley
repopulation effect since for an {\em unrelaxed} structure of an
{\em ideal} Si crystal, the results obtained by Wilson and Feher
\cite{wilson61} are nicely reproduced after the self-consistent
cycle (cf. open squares in Fig.~\ref{fig:sum}).

We are now able to take into account explicitely {\em by first
principles} the strain of the Si lattice as well as the relaxation
around the P donors within. For the optimization of the strained
Si cells, we used the efficient self-consistent charge DFT-based
tight binding (SCC-DFTB) approach~\cite{frauenheim00}.
Optimisation of long slabs (up to 12 unit cells along the [001]
direction) show an almost linear dependence of the compression
along [001] as an answer to the tensile strain in the (001)-plane,
effectively following linear elasticity theory as indicated in the
inset of Fig.~\ref{fig:sum}. This result confirms that {\em
linear} elasticity theory remains valid in the complete regime
investigated, even up to pure germanium as a substrate
($\chi\approx$\,\,-89). The hf parameters calculated with LMTO-GF
under these assumptions (cf. open circles in Fig.~\ref{fig:sum})
already become smaller since the donor wavefunction becomes more
delocalized as a result of the enhanced volume, exceeding the
high-stress limit of $0.33\cdot A_{\rm{HF}}$ obtained above.

This tendency is enhanced, if local relaxation around the P donors
is taken into account: According to SCC-DFTB calculations on
large, explicitly strained supercells with 512 atoms, this
relaxation is dominated by a slight reduction of the bond-length
between the P donor and its nearest Si ligands by about 1\%,
nearly independent of the strength of the tensile strain in the
plane of the Si epilayer and already present in the unstrained
case. Re-calculating $A_{\rm HF}(\chi)$ for this geometry with the
LMTO-GF code, we find a further reduction (full triangles in
Fig.~\ref{fig:sum}).
While the theoretically predicted absolute value $A_{\rm HF}$ is
$\approx 2$~mT too large in the unstrained case without nearest
neighbour relaxation, $A_{\rm HF}$ is now in good accordance with
the experiment for all $\chi$.
The hf interaction observed experimentally in the moderately
strained P-doped Si layers can, thus, be explained by the
increased volume of the unit cell together with a slight inward
relaxation of the nearest Si neighbors. Since already such a small
relaxation has a huge influence on the predicted relative hf
splittings for the strained material, the remaining discrepancy
between experiment and theory can easily be attributed to
uncertainties due to the well-known flatness of the total energy
surface in Si \cite{gerstmann02}. According to our DFT
calculation, the decrease of the central P-related hf interaction
is accompagnied by a remarkable increase of the superhyperfine
interaction with neighbouring $^{29}$Si atoms. Comparative
electron-nuclear double resonance (ENDOR) or electrically detected
ENDOR (EDENDOR) \cite{stich96} measurements could be used to
verify these predictions.

Finally, we briefly turn to the anisotropy of the $g$-factor of
the donor. The values observed here are somewhat larger than the
values
reported by Wilson and Feher~\cite{wilson61} and those observed in
strained two-dimensional electron gases~\cite{graeff99}. Note that
a single-valley effect arising from an admixture of the doublet is
not expected for uniaxial strain in [001] direction, dominant in
our samples. Rather, a pure repopulation effect is
expected~\cite{wilson61, roth60, liu61}. However, a detailed
calculation of the $g$-factor under strain including local
relaxation appears warranted to understand the origin of the
larger $\Delta g$ observed in our samples.

To summarize, we have presented an experimental and theoretical
study of the hyperfine splitting of phosphorus donors in strained
layers up to high strain levels of $\epsilon_\perp$=$-0.00882$.
The splitting is reduced to values far below predictions published
so far. Density functional theory demonstrates that repopulation
of the CB minima by strain, the change of the unit cell volume,
and a relaxation of the bond length between the P donors and the
next nearest Si neighbours are required to account for the
observed hyperfine interaction. Our results indicate that there
exists no high-stress limit for the reduction of the P-related hf
splitting.

This work was supported by the Deutsche Forschungsgemeinschaft (SFB
631).


\newpage
\end{document}